%% file: 0main.tex
\documentclass[sigconf]{acmart}

\usepackage{listings}

\usepackage{color}
\usepackage{colortbl} 
\usepackage{xcolor}
\usepackage{xspace}
\usepackage{algorithm}
\usepackage{algorithmicx}

\usepackage{xcolor} % For custom colors

\definecolor{codegreen}{rgb}{0,0.6,0}
\definecolor{codegray}{rgb}{0.5,0.5,0.5}
\definecolor{codepurple}{rgb}{0.58,0,0.82}
\definecolor{backcolour}{rgb}{0.95,0.95,0.92}

\AtBeginDocument{%
  \providecommand\BibTeX{{%
    \normalfont B\kern-0.5em{\scshape i\kern-0.25em b}\kern-0.8em\TeX}}}

\setcopyright{acmlicensed}
\copyrightyear{2018}
\acmYear{2018}
\acmDOI{XXXXXXX.XXXXXXX}

\acmConference[Conference acronym 'XX]{Make sure to enter the correct
  conference title from your rights confirmation emai}{June 03--05,
  2018}{Woodstock, NY}
\acmISBN{978-1-4503-XXXX-X/18/06}

\newcommand{\datasetFont}{\text}
\newcommand{\ours}{\datasetFont{CityFlowER}\xspace}

\begin{document}

\title{\ours: An Efficient and Realistic Traffic Simulator with  Embedded Machine Learning Models}

\author{Longchao Da$^1$, Chen Chu$^1$, Weinan Zhang$^2$,  Hua Wei$^1$}
\affiliation{%
  \institution{$^1$Arizona State University, $^2$Shanghai Jiao Tong University\\
  {$^1$\{longchao, cchu37, hua.wei\}@asu.edu, $^2$wnzhang@sjtu.edu.cn}}
  \city{Tempe, AZ}
  \country{USA}
}

\renewcommand{\shortauthors}{Trovato and Tobin, et al.}

\begin{abstract}
Traffic simulation is an essential tool for transportation infrastructure planning, intelligent traffic control policy learning, and traffic flow analysis. Its effectiveness relies heavily on the realism of the simulators used. Traditional traffic simulators, such as SUMO and CityFlow, are often limited by their reliance on rule-based models with hyperparameters that oversimplify driving behaviors, resulting in unrealistic simulations. To enhance realism, some simulators have provided Application Programming Interfaces (APIs) to interact with Machine Learning (ML) models, which learn from observed data and offer more sophisticated driving behavior models. However, this approach faces challenges in scalability and time efficiency as vehicle numbers increase.
Addressing these limitations, we introduce \ours, an advancement over the existing CityFlow simulator, designed for efficient and realistic city-wide traffic simulation. \ours innovatively pre-embeds ML models within the simulator, eliminating the need for external API interactions and enabling faster data computation. This approach allows for a blend of rule-based and ML behavior models for individual vehicles, offering unparalleled flexibility and efficiency, particularly in large-scale simulations. We provide detailed comparisons with existing simulators, implementation insights, and comprehensive experiments to demonstrate \ours's superiority in terms of realism, efficiency, and adaptability. 
\end{abstract}


\maketitle

\input{1intro}

\input{2related}
\input{3method2}

\input{4perform}

\input{5demo}

\section{Summary}

This work presents a novel way to extend and enhance the existing simulator for realistic simulation and high efficiency by embedding the behavior models into the acting functions within the source architecture of the simulator. It provides flexible behavior model support and avoids unnecessary data transmission between the simulator and the Python interface. By constructing \ours, we verified the feasibility of the proposed method on a renowned base simulator CityFlow. We transplant two learned models (\textit{followSpeed} and \textit{laneChange}) into the simulator, and showcase how similarly the vehicles in the new simulator can perform as expected. The code will be released 
 and actively maintained in public repository\footnote{\url{https://github.com/cityflow-project/CityFlowER.git}}.

\section*{Acknowledgments}
We appreciate the support from CityFlow team, including Huichu Zhang and Zhenhui (Jessie) Li for their suggestions during the development of this paper. The features enabled in this paper will be further integrated into CityFlow code base as new versions.

\bibliographystyle{ACM-Reference-Format}
\bibliography{sample-base}

% %%
% %% If your work has an appendix, this is the place to put it.
% \appendix









\end{document}

%% file: 1intro.tex
\section{Introduction}
% 1. Traffic Simulation is important (transportation infrastructure planning, intelligent traffic signal control policies, traffic flow analysis, etc)
Traffic simulation plays a crucial role in analyzing potential problems~\cite{ameli2020simulation}, and provides quantitative results for decision-making~\cite{zhang2019cityflow} in transportation systems. %In the era of consistently growing human civilization, traffic simulation is an effective way to boost development progress, such as providing transportation infrastructure planning, intelligent traffic signal control policies verification, and traffic flow pattern analysis, etc. % 2. Current traffic simulator as a solution. (pros and cons), and how to make it more realistic. 
In essence, the insights coming from simulators have a prerequisite: the simulator is realistic enough to replicate real-world traffic problems. With simulators being unrealistic, the analysis and solutions coming from the simulators will be unreliable.

\begin{figure*}[t!]
    \centering
    \includegraphics[width=0.90\linewidth]{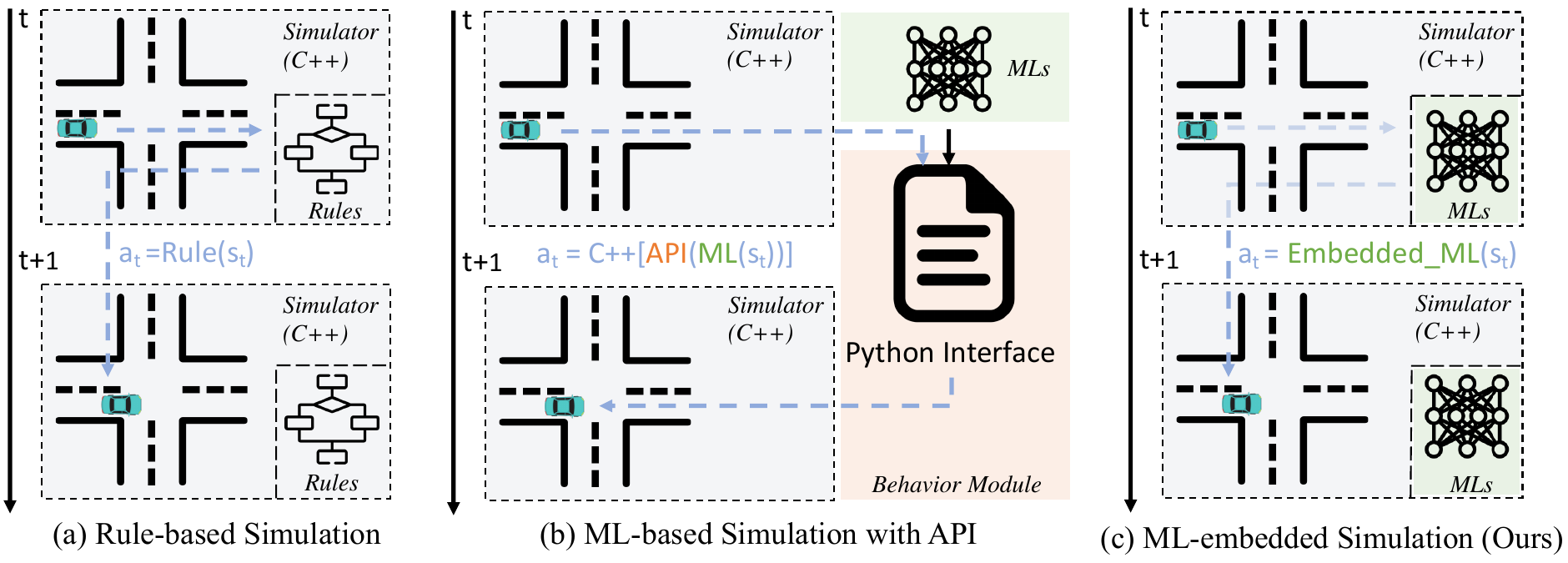}
    \vspace{-5mm}
    \caption{Comparison on \ours of embedding ML models \textit{vs} rule-based method and interaction-based method in original CityFlow for realistic simulations. The original simulator is implemented in C++ for efficiency. Note that the $\textit{MLs}$ can be any vehicle behavior models, e.g., \texttt{Car Following Model}, \texttt{Lane Changing Model}, etc.}
    \label{fig:framework}
\end{figure*}

In order to make the simulators realistic, existing work usually calibrates the hyperparameters on the given physical/rule-based models within the simulator~\cite{tsai2020optimizing, sommer2019veins,RONDINONE201399,ramamohanarao2016smarts}. Traffic simulators, especially microscopic simulators like VISSIM~\cite{fellendorf1994vissim}, SUMO~\cite{behrisch2011sumo} or CityFlow~\cite{zhang2019cityflow}, have different rule-based models within, as shown in Figure~\ref{fig:framework} (a),  given the current state $s_t$, the action $a_t$ will be output from rule-based models. Representative models include car-following models, lane-changing models, or routing models, each of which consists of multiple hyperparameters and assumes that the behavior of the vehicle
is only influenced by a small number of factors with predefined rule-based relations, such as Wiedemann and Krauss, etc. Calibrating the hyperparameters for these models will need to find the parameters that best fit the observed data~\cite{shahriari2020ensemble}. The problem with such methods is that their assumptions oversimplify the driving behavior, resulting in the simulated driving behavior far from the real world.

To make the behavior model more realistic, some simulators provide Application Programming Interfaces (APIs) to control the vehicle behaviors~\cite{that2011integrated, jeihani2017integration}. These APIs extract vehicle states from the simulators, upon which the ML models outside of the simulators take as input and output the predictions back into simulators through the APIs, as illustrated in Figure~\ref{fig:framework} (b). These ML models do not assume the underlying form of the behavior models and directly learn from the observed data~\cite{feng2023trafficgen,suo2021trafficsim, egea2019vehicular, shi2023development}. With ML models, a more sophisticated driving behavior model can be represented by a parameterized model like neural nets and provides a promising way to learn the models that behave similarly to observed data. This form of behavior API provides flexibility in vehicle behavior since different ML models can be utilized to interact with the simulator. However, as the number of vehicles scales up, the interaction between the API and ML models outside the simulators requires additional time consumption by extra computation and data transition, where a simulator must be fast enough for decision-making.

% Current traffic simulation relies on sophisticated simulators, and the quality of the simulators is the bottleneck for various downstream tasks. Building a simulator is usually an arduous project that takes much time and energy, and the software architecture design normally determines its practicability and extensibility. If the development of the simulator can not provide expansibility that supports the transplant of ever-changing technologies, much effort would be wasted. For traffic simulators, most classic ones still rely on hardcode or rule-based behavior models that control the vehicles' movements, these methods face realistic problems that can hardly provide meaningful insights due to the gap between simulators and the real world. There are some explorations that tries to apply Machine Learning methods to learn behavior models and use the generated actions as the vehicles' actions, which provides useful and customizable action-making, bringing the simulator closer to realism. However, the traditional way of using the trained model's action-taking is based on the interaction between the simulator's vehicle agents and the behavior module's model interfaces, which introduces much more unnecessary time consumption.
To enable \textbf{\underline{E}}fficient and \textbf{\underline{R}}ealistic simulation for intelligent transportation, we present CityFlow\textbf{ER} 
based on a well-accepted traffic simulator CityFlow~\cite{zhang2019cityflow} to support the realistic city-wide traffic simulation. Specifically, this work pre-embeds ML models into the simulator and controls vehicle behaviors directly within the simulator. Instead of relying on interactions between APIs and ML models outside of the simulator, \ours directly conducts data computation in the simulator with the embedded ML models to enable efficient simulation. Additionally, in \ours, any individual vehicle can be specified with either rule-based behavior models or ML models, whereas original CityFlow or SUMO only supports rule-based models. This ML-embedded mechanism provides diverse and flexible simulations while maintaining high efficiency for large-scale traffic simulation. We have elaborated the proposed method in comparison to the existing simulators for realistic simulation, and we also provide implementation details and conduct sufficient experiments to show the correctness, efficiency, and flexibility in \ours's simulation.

%% file: 3method2.tex
\section{DESCRIPTION}

In this section, we first propose a simulator-level behavior model pre-embedding mechanism, and then provide an implementation in a renowned simulator CityFlow~\cite{zhang2019cityflow} to verify the feasibility and efficiency.

\subsection{Behavior Model Pre-embedding Mechanism}

We propose a way to embed the pre-trained model directly into the simulator itself and avoid extra simulator-model communication. As shown in Figure~\ref{fig:framework}(c), ML models are pre-loaded inside of the simulator in correspondence to the acting functions, and compiled readily for use, given the current environment state $s_t$, the appropriate action $a_t$ can be decided based on the model's logics while keeping the simulation efficiency, where $a_t := (v_t, l_t)$, $v_t$ represents the speed adjustment that $v_t \in \mathbb{R}$ and $l_t$ represents lane changing options: $l_t \in \{\text{left}, \text{stay}, \text{right}\}$. By pre-embedding the models, the interaction between the simulator and ML models is avoided, thus data flow time overhead is greatly reduced.

\subsection{System Implementation in \ours}
In order to implement an extensible, realistic, and highly efficient simulator, this work builds upon the existing simulator CityFlow~\cite{zhang2019cityflow}, aiming at improving the realistic behavior modeling while keeping simulating efficiency. We follow the same basic \texttt{Road Network} structure like \textbf{Segments} as defined in CityFlow for its computation efficiency. For \texttt{Car Following Model} and \texttt{Lane Changing Model}, we implemented the embedded architecture as shown in Figure~\ref{fig:framework} (c), we leverage the LibTorch~\cite{paszke2019pytorch}, a C++ supportive release of Pytorch to realize the model loading, an example of $\textit{Vehicle::getCarFollowSpeed}$ is shown below. In this implementation, the ML $speedModel$ is loaded with the specified $\textit{PATH}$ at the start of the simulation. At each time step $t$, the $\textit{input\_feature}$ is the observable state information $s_t$, fed into the loaded $speedModel$, an output from $speedModel$ is the $carFollowSpeed$ expected to take in the next step.
% \begin{center}
% \begin{minipage}{.95\textwidth}
% \label{lis:code}
% \begin{lstlisting}[language=Python, caption=The Class and API Format , label=code:api-example]
% class AugmentTask:
%     def __init__(self, params) -> None:
%         self.params = params
%         # set possible pre-defined params, e.g., folder path

%     @func_prompt(name="prompt name",
%         description="""detailed explanation""")

%     def embody(self, target: str) -> str:
%         try:
%             result = Execution(target)
%             # Concrete Execution Implementation
%         except Error as e:
%         print(e)
%         return result
            
% \end{lstlisting}
% \end{minipage}
% \end{center}

\begin{lstlisting}[language=C++]
// vehicleInfo is updated out of the function
// "PATH" is the pre-trained model path
double Vehicle::getCarFollowSpeed(double interval) {

    Vehicle *leader = getLeader();
    std::vector<torch::jit::IValue> input_feature;
    auto input = torch::tensor ({leader->getSpeed(), vehicleInfo.speed ...});

    input_feature.push_back(input);
    static torch::jit::script::Module speedModel = torch::jit::load("PATH");

    at::Tensor output_speed = speedModel.forward(input_feature).toTensor();
    return output_speed.item<double>();
}
\end{lstlisting}

Similarly, the lane-changing model is an ML model implemented to generate the lane choice: \{0: "change to the left (Inner lane)", 1: "keep current lane", 2: "change to the right (Outer lane)"\}. The ML model is also loaded at the beginning of the simulation. 
% \begin{minted}[frame=lines, framesep=2mm, baselinestretch=1.2, fontsize=\footnotesize]{c++}
% void SimpleLaneChange::makeSignal(double interval) {

% std::vector<torch::jit::IValue> input_feature;
% auto input = torch::tensor ({vehicle->engine->getCurrentTime(),...});
% input_feature.push_back(input);

% static torch::jit::script::Module laneModel = torch::jit::load("PATH");
% at::Tensor lane_choice = laneModel.forward(input_feature).toTensor();
% signalSend->target = lane_choice;
% }
% \end{minted}

\begin{lstlisting}[language=C++]
void SimpleLaneChange::makeSignal(double interval) {

    std::vector<torch::jit::IValue> input_feature;
    auto input = torch::tensor ({vehicle->engine->getCurrentTime(),...});
    input_feature.push_back(input);

    static torch::jit::script::Module laneModel = torch::jit::load("PATH");
    at::Tensor lane_choice = laneModel.forward(input_feature).toTensor();
    signalSend->target = lane_choice;
}
\end{lstlisting}

In \ours, the realistic simulation relies on realistic behavior models, which are trained by behavior data using Imitation Learning (IL) or Behavior Cloning (BC) methods. The behavior data could either be collected from a more realistic simulator's rollout process or from real-world demonstrations. 

It is worth noting that, to guarantee the universality across different simulators, defining a unified input feature space and output action space is inevitable. CityFlow is rich in system state information APIs and provides a full description of microcosmic vehicle properties, so based on CityFlow perceptible features, we define a superset $F_{\phi} = \{f_1, f_2, f_3, ... f_n\}$, $\phi = \{\textit{followSpeed, laneChange}\}$, and the feature set $G$ from behavior dataset can be represented as $G \subseteq F$. If for the trained model that adopts feature $G$, exists $G \subset F$, then when applying CityFlow observed features, the $k$ extra features: $\{f_{e^1}, f_{e^2}, ..., f_{e^k}\}$ will be masked to align the feature space. For output space, the output of $\textit{followSpeed}$ is the real speed value, and the output of $\textit{laneChange}$ is the lane-changing choice.

Once the system has access to behavior models,  the simulator will be rebuilt based on the new structure with embedded models. In execution, these models will control corresponding behaviors.

% \begin{figure}[h!]
%     \centering
%     \includegraphics[width=0.95\linewidth]{figs/demo.pdf}
%     \caption{The overview of the \ours for more realistic behavior modeling}
%     \label{fig:levels}
% \end{figure}

% \subsection{Incorporate the Customizable Models}

% The Python Interface is designed as below (\textcolor{red}{to be finished}):...

% model requirement

% \textcolor{red}{how to load the ML models into the simulator}

% \begin{center}
% \begin{minipage}{.95\textwidth}
% \label{lis:code}
% \begin{lstlisting}[language=Python,  label=code:api-example]
% def get_following_speed(params):
%     try:
%         result = Execution(target)
%     except Error as e:
%     print(e)
%     return result
            
% \end{lstlisting}
% \end{minipage}
% \end{center}

% \subsection{Front End}

%% file: 4perform.tex
\section{Performance}

In this section, we conduct experiments to verify the accuracy, efficiency, and flexibility of the proposed method.

\subsection{Accuracy}
\paragraph{\ours to recover the CityFlow driving behaviors}\label{sec:citycity}

\begin{figure}[h!]
    \centering
    \begin{tabular}{cc}
     \includegraphics[width=0.93\linewidth]{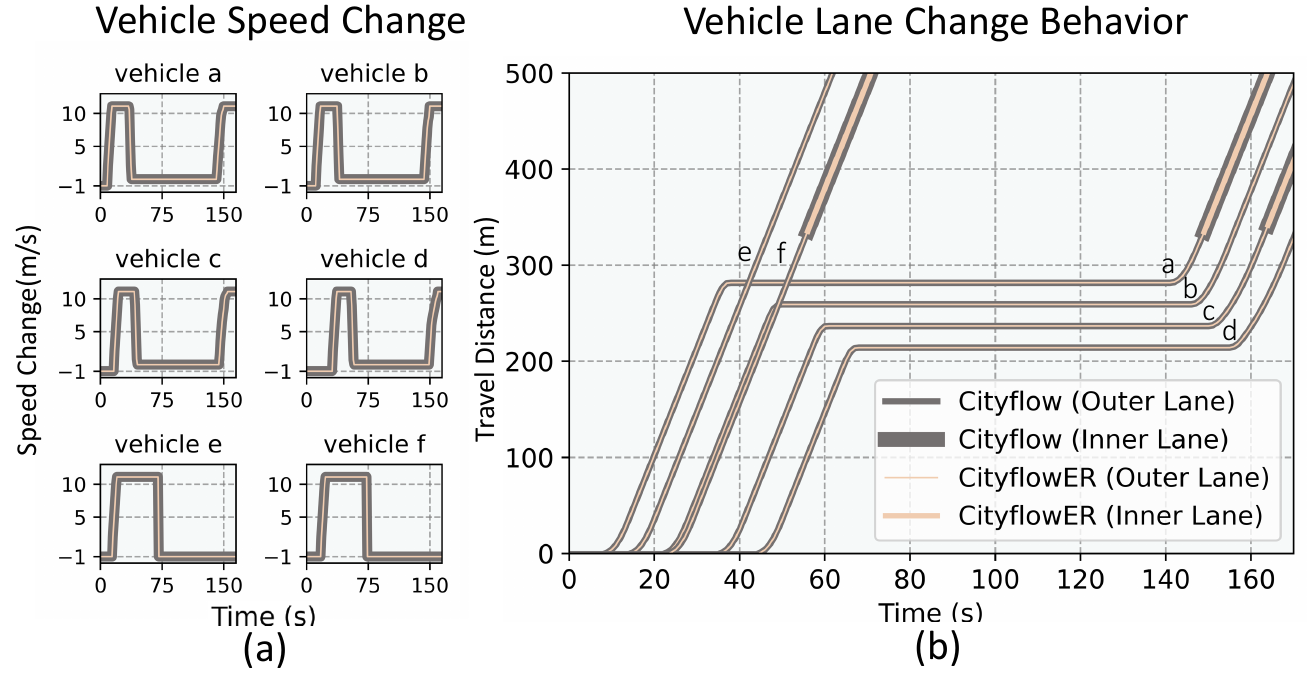} 
     % \includegraphics[width=0.45\linewidth]{figs/cityflowErrors.png}\\
      % Speed Comparison  
      % (a) Speed Difference & (b) Trajectory Errors    
    \end{tabular}
    \vspace{-3mm}
    \caption{The vehicles behavior comparison of \ours-pink and Cityflow-grey on example vehicles (a$\sim$f). Left: the speed changes over time. Right: The lane changes over time.}
    \label{fig:citycity}
\end{figure}

In this experiment, we investigate if \ours could recover the original CityFlow driving behaviors, i.e., car-following speeds and lane-changing behaviors. We trained two behavior models by BC using data logged from CityFlow in network hz1x1. In testing, we visualize speed change in Figure ~\ref{fig:citycity} (a). In this sub-figure, we could observe the same speed changes for different vehicles in \ours and CityFlow, which indicates an accurate imitation of car-following behavior. In Figure ~\ref{fig:citycity} (b), we visualize the vehicle trajectories related to the lane-changing behaviors, and found out that the trajectories of the same vehicle are almost identical between CityFlow and \ours, indicating the ML models in \ours could recover the rule-based vehicle behavior models in original CityFlow.

% which we quantitatively evaluate from the trajectory error aspect, we adopt Average Displacement Error (ADE) and Final Displacement Error (FDE), and as shown in sub-figure (b), the low trajectory errors also show reliable lane changing behavior recovery abilities.

\paragraph{\ours to recover the SUMO driving behaviors}

In Section~\ref{sec:citycity}, we have shown that \ours is able to recover the CityFlow's behavior models accurately, however, the original CityFlow can only support a singular behavior model, while \ours is designed in a broader way, so we designed an experiment to compare the \ours with learned behavior models (using data logged from SUMO) to original SUMO, the models are commonly trained with simple MLP layers for 300 Epochs as in Section~\ref{sec:citycity}, we have testified the total 6 combinations of settings\footnote{\url{https://sumo.dlr.de/docs/Definition_of_Vehicles,_Vehicle_Types,_and_Routes.html}} (3 car-following models: Krauss, IDM, and BKerner, and 2 lane-changing models: LC2013, and SL2015) as shown in Table~\ref{tab:citysumo}, we conduct simulations in both SUMO and \ours using the approximately identical map and with same 122 vehicle configs\footnote{\url{https://github.com/DaRL-LibSignal/LibSignal/tree/master/data/raw_data}} as released in LibSigsnal~\cite{mei2023libsignal}. We calculate the mean of vehicles' final displacement error and travel time consumption error. As shown in Table~\ref{tab:citysumo}. Just by simple training, it can achieve low errors, indicating the potential to replicate realistic behaviors using models learned from real-world data.

\begin{table}[t]
\small
\centering
\caption{The Ability to recover different behavior models from other simulators (SUMO). The behavior models are a combination of trained models for car-following and lane-changing behavior using data collected from SUMO.}
% , V1: roads with lighter loaded vehicles, V2:roads with heavier loaded vehicles, V3: roads under rainy weather, V4: roads under snowy weather
\label{tab:citysumo}
\setlength{\tabcolsep}{1mm}
    \begin{tabular}{cccccc}
    \toprule
    Behavior Models  & Final Displacement Error & Travel Time Error      \\ \midrule
    % v1 ==> v3, v2 ==> v4
    Krauss + SL2015 & 1.24$_{\pm\text{0.71}}$ & 0.44$_{\pm\text{0.26}}$ \\
    Krauss + LC2013 & 0.64$_{\pm\text{0.52}}$ & 0.30$_{\pm\text{0.09}}$  \\

     \midrule
    IDM + SL2015   &   1.82$_{\pm\text{1.64}}$ & 0.56$_{\pm\text{0.91}}$ \\
     IDM + LC2013   &   0.79$_{\pm\text{0.29}}$ & 0.50$_{\pm\text{0.37}}$ \\

     \midrule

    BKerner + SL2015   &  1.26$_{\pm\text{0.35}}$ & 0.75$_{\pm\text{0.16}}$  \\
    BKerner + LC2013   &  0.29$_{\pm\text{1.13}}$ & 0.08$_{\pm\text{0.21}}$  \\
    \bottomrule
    \end{tabular}
\end{table}

\subsection{Efficiency}

\begin{figure}[h!]
    \centering
    \includegraphics[width=0.80\linewidth]{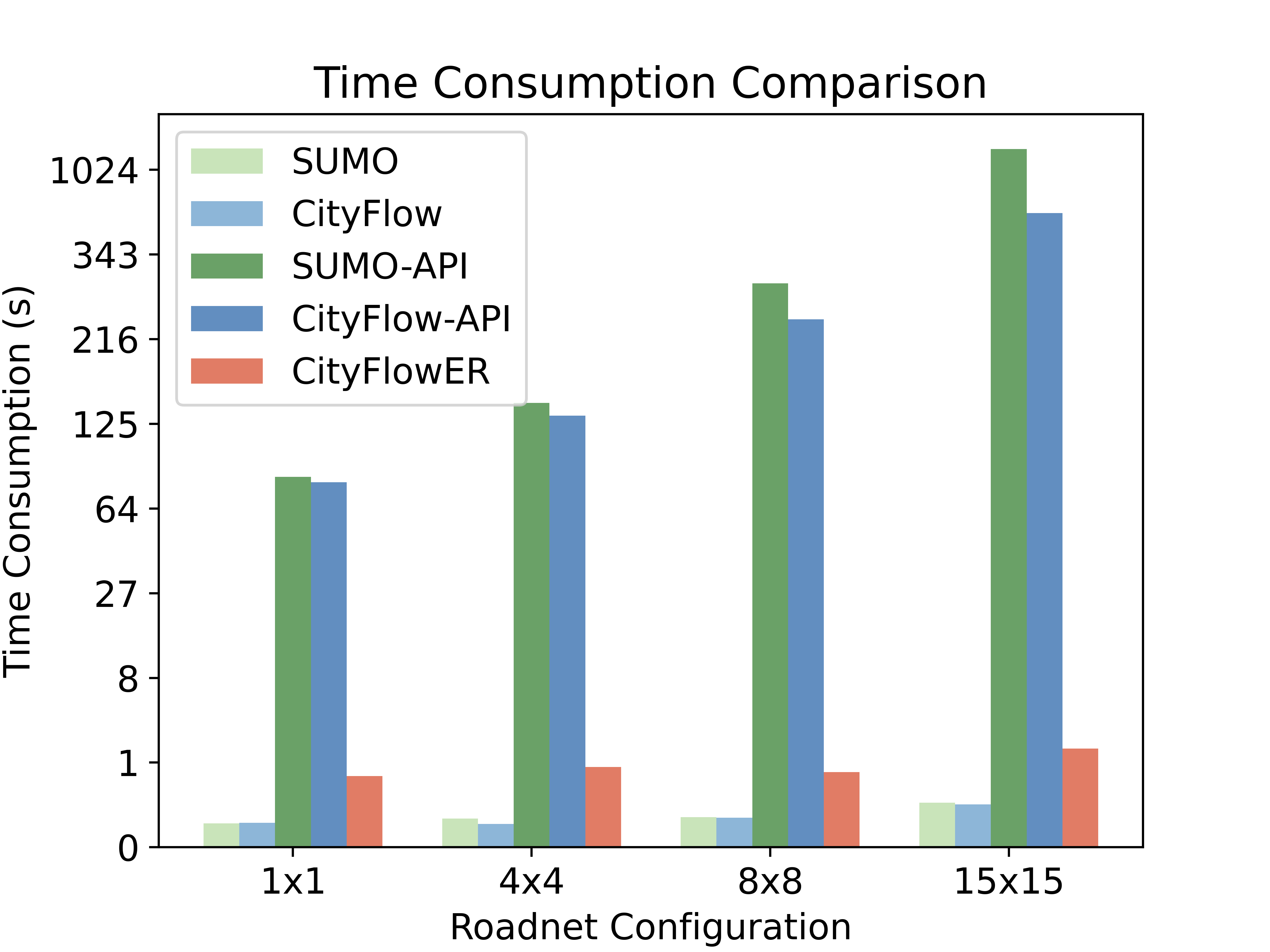}
    \caption{Speedup of SUMO, CityFlow and \ours}
    \label{fig:speedup}
\end{figure}

In this section, we investigate the efficiency of the \ours compared to baseline simulations: SUMO, Cityflow-API (integrate the behavior models by commonly adopted API controls), and SUMO-API.  We could observe that the API-based realistic simulation suffers from high time consumption due to the expensive behavior module to simulator communications, while the \ours could provide minimal efficiency impairment compared to CityFlow. The comparison metric is the time consumption in seconds. As shown in Figure~\ref{fig:speedup}, \ours achieves similar efficiency to rule-based simulators and is much faster than the API-based simulations.

\subsection{Diversity and Flexibility}
\begin{figure}[h!]
    \centering
    \includegraphics[width=0.95\linewidth]{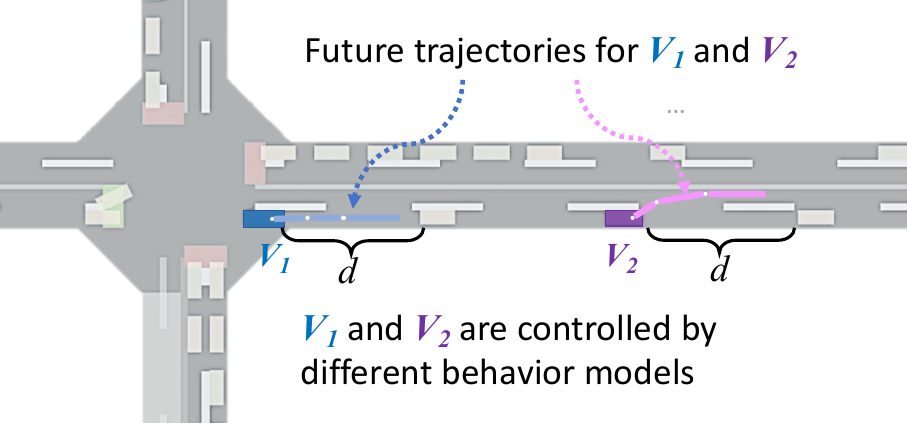}
    \caption{The diverse behaviors of vehicles controlled by different models in \ours. The figure shows a snapshot of traffic flow at time $t$. Since $V_1$ and $V_2$ are specified with two different behavior policies, they behave differently in the following time steps even though under a similar driving situation (distance with lead vehicle, moving velocity, and lane capacity).}
    \label{fig:behaviors}
\end{figure}

\ours can support specifying individual driving behavior models for individual vehicles, which provides the simulating of diverse behaviors from different ML models and flexibility of specifying different behaviors for different vehicles.
As shown in Figure~\ref{fig:behaviors}. Applying different control policies could help the simulator to customize the driving models on individual vehicles and achieve a realistic simulation with the idea that each driver might have their own driving behavior in the real world. This would provide more insightful research results and reliable traffic planning strategies.

%% file: 5demo.tex
\section{DEMO details}
We provide a video\footnote{\url{https://drive.google.com/drive/folders/1O6-HR8HgNoMEBzqAJWpRz5p6-3l7VtPy?usp=sharing}} demonstration of \ours for following scenarios to show its capability for realistic and efficient simulation. 
\begin{itemize}
    \item Use \ours to recover the original CityFlow.
    \item Use \ours to recover the original SUMO behavior.
    \item Enable Diverse behaviors for individual vehicles. 
    \item Conduct large-scale (30x30 intersections) efficient simulation with diverse behaviors.
    
\end{itemize}